\documentclass[twocolumn]{article}
\usepackage[utf8]{inputenc}
\usepackage{graphicx}
\usepackage{authblk}
\usepackage{caption}
\usepackage[numbers]{natbib}
\usepackage{amsmath}
\usepackage{amssymb}
\usepackage[a4paper,margin=1.5cm,footskip=1cm]{geometry}

\title{Spontaneous domain formation in disordered copolymers as a mechanism for chromosome structuring}

\author[1]{Matteo Negri}
\author[1]{Marco Gherardi}
\author[2]{Guido Tiana}
\author[3]{Marco Cosentino Lagomarsino}
\affil[1]{Department of Physics, Universit\'a degli Studi di Milano, via Celoria 16, 20133 Milano, Italy}
\affil[2]{Center for Complexity and Biosystems and Department of Physics, Universit\'a degli Studi di Milano and INFN, via Celoria 16, 20133 Milano, Italy}
\affil[3]{Sorbonne Universites, UPMC Univ Paris 06, UMR 7238, Computational and Quantitative Biology, 4 Place Jussieu, Paris, France; CNRS, UMR 7238, Paris, France; IFOM, Milan, Italy}

\begin{document}

\maketitle

\begin{abstract}
  Motivated by the problem of domain formation in chromosomes, we
  studied a co--polymer model where only a subset of the monomers feel
  attractive interactions. These monomers are displaced randomly from
  a regularly-spaced pattern, thus introducing some quenched disorder
  in the system.
  Previous work has shown that in the case of regularly-spaced
  interacting monomers this chain can fold into structures
  characterized by multiple distinct domains of consecutive
  segments. In each domain, attractive interactions are balanced by
  the entropy cost of forming loops. We show by advanced
  replica-exchange simulations that adding disorder in the position of
  the interacting monomers further stabilizes these domains.
  The model suggests that the partitioning of the chain into
  well-defined domains of consecutive monomers is a spontaneous
  property of heteropolymers. In the case of chromosomes, evolution
  could have acted on the spacing of interacting monomers to modulate
  in a simple way the underlying domains for functional reasons.
\end{abstract}

\section{Introduction}

Simple heteropolymer models provide a candidate explanation for the
formation of intermediate- and large-scale domains in prokaryotic and
eukaryotic
chromatin~\cite{imakaev2015modeling,Lagomarsino2015,Nicodemi2014,Dame2011,schwarzer2017two}.
Such domains may be defined as extended contiguous regions along the
DNA chain in which the DNA interacts preferentially with sites of the
same domain. As such, they appear as squared blocks in the contact
matrix of the polymer, which is measurable by chromosome capture and
sequencing techniques~\cite{Nicodemi2014}.
While other mechanisms, such as loop
extrusion~\cite{Fudenberg2016,Goloborodko2016a} likely contribute to
driving domain formation, the interaction between chromosome-bound
proteins is considered to be one of the main drivers for this
behavior.
For example, in mammals, the protein CTCF has been shown to form dimers \cite{pant2004mutation} that can stabilize chromatin loops.
In bacteria, the proteins H-NS and
MatP have the same bridging capabilities~\cite{Dame2011,RN49}.
One main question is what drives domain identity, size and stability,
and to what extent intra-specific interactions are needed to form
domains. In other words, while it is reasonable to think that the
domain formation is mediated by proteins that are bound to chromatin
and that interact with each other, we do not know how many species are
needed to program a certain number of domains into a
polymer~\cite{Nicodemi2014}.  Since there are thousands of domains at different scales in
mammalian genomes, trivially associating one--to--one interactions would require the presence of thousands of different types of
intra-specific DNA-binding proteins. It is more reasonable to think
that only a small number of proteins is responsible for the
interactions between the chromatin sites.

Focusing on the direct interaction between chromatin structure
factors, various kinds of heteropolymer
models~\cite{Nicodemi2014,brackley2016simulated,
  nazarov2015statistical,junier2010spatial} have been proposed, to
explain various aspects of domain formation, specification and
stability. Perhaps the simplest one is a polymer chain in which
equally-spaced monomers attract monomers of the same
type~\cite{junier2010spatial,scolari2015combined}. This is a specific
type of co--polymer model in which only one of the two chemical
species exerts attractive interactions (and the linear density of this
species is typically considered to be low). This model shows that
multiple-domain states are possible without any intra-specific
interaction~\cite{scolari2015combined}.  In such states, the polymer
is collapsed into a multiple rosette configuration. Analytical
arguments support the hypothesis that such multi-domain phase is
stable, and due to the trade-off between the surface-tension cost of
keeping a core of bridging proteins and the entropy cost of the arms
of the rosette states.

Here, we use replica-exchange Monte Carlo (MC) simulations to explore
the equlibrium states of the disordered version of this model, where
the interacting monomers are not equally spaced, but arranged randomly
along the backbone in a fixed (quenched) configuration. We ask about
the role played by these disordered interactions into the
thermodynamic stability of the collapsed states with one and multiple
domains. We also address the possible role of the disorder into
localizing the domains in a specific region of the chain, which may
lead to pre-programmed spatial domains without intra-specific
interactions.

\section{Model}

\begin{figure} 
\centering
\includegraphics[width=0.5\textwidth]{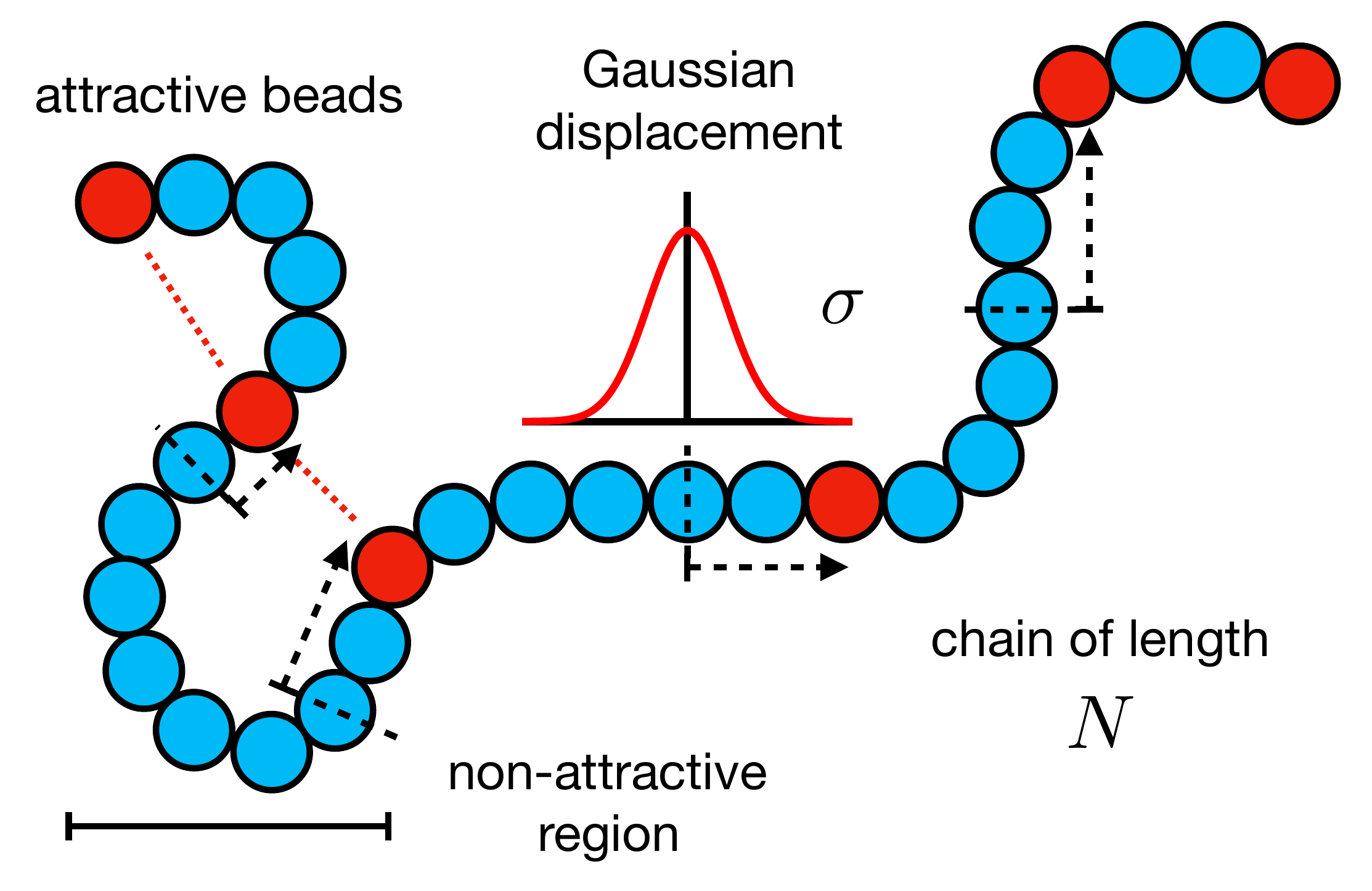}
\caption{Sketch of the model used in this work. The polymer is made of
  two types of monomers. Short-ranged attractive monomers (red) are
  separated by regions of non-attractive ones (light-blue). The
  position of attractive monomers is fixed at distance extracted from
  a Gaussian distribution, and attractive monomers are fixed during
  each simulation (quenched disorder). Monomers are described as
  hard-sphere beads, joint by inextensible links.}
\label{fig:model}
\end{figure}

We study a coarse-grained model consisting in a polymer made of $N$
consecutive monomers represented as hard-sphere beads of radius
$R_{\mathrm{HC}}$ (see Fig.\ref{fig:model}).  Each monomer
represents a region of the chromosome, and the size can be defined at will to describe the fiber at any resolution (e.g., from the finest experimental resolution of $\sim$ kb to describe topological associating domains, to that of Mb to describe chromosomal compartments).
In this model, bead $i$ can interact with bead $j$ with an attractive
short-ranged square well potential $u_{ij}$:
\begin{displaymath}
  u_{ij} = \begin{cases} \infty & \mbox{if } r_{ij} < R_{\mathrm{HC}} \\
    B_{ij} & \mbox{if } R_{\mathrm{HC}} < r_{ij} < R \\ 0 & \mbox{if }
    r_{ij}>R \ , \end{cases}
\end{displaymath}
where $r_{ij}$ is the distance between the beads, $R_{\mathrm{HC}}$ is
the hard-core radius, $R$ is the range of the interaction and $B_{ij}$
is the interaction energy, which depends on the types of the monomers
$i$ and $j$. In order to represent bridging interactions, we place $p$
attractive monomers along the chain (see Fig.\ref{fig:model}). Therefore, the interaction energy is
\begin{displaymath}
  B_{ij} = \begin{cases} -\varepsilon & \mbox{if } i \mbox{ and } j
    \mbox{ are attractive monomers} \\ 0 & \mbox{otherwise} \end{cases}  
\end{displaymath}
where $\varepsilon>0$ since the interaction between bridging points is
always attractive.

Using square-well potentials makes the MC calculations easier and
faster than using smooth short-ranged potential. The uncrossability of
the polymer chain is guaranteed by the hard-core repulsion, whose
range is $R_{\mathrm{HC}}=0.472\lambda$. The distance $\lambda$
between consecutive beads is maintained fixed by the MC moves, and
sets the microscopic length scale, with respect to which all the
lengths of the model are measured.

We first studied regular co--polymers, in which interacting monomers are places every $p$ other monomers which only repel each other by hard--core repulsion. Subsequently, we studied a disordered model in which the position of these interacting monomers is displaced by a Gauassian--distributed quantity.

The simulations are performed with an off-lattice MC algorithm whose degrees of freedom are the angles and dihedrals of the chain, updated with flip and pivot moves through a Metropolis acceptance rule, to ensure an effective sampling of the canonical ensemble. The algorithm is implemented in a freely-distributed code \cite{tiana2015montegrappa}.
To improve the efficiency of the algorithm to sample equilibrium conformations also at low temperatures, the MC algorithm is used in its parallel-tempering variant, in which 16 replicas of the system are simulated in parallel at increasing temperature, and the conformations of adjacent temperatures are exchanged every 1000 MC step with a Metropolis-like acceptance rule \cite{swendsen1986replica}. The thermodynamic quantities are then calculated with a weighted--histogram technique \cite{ferrenberg1989optimized}.

\section{Results}
\subsection*{Multi-domain states in absence of disorder are
  stable}

\begin{figure}
\centering
\includegraphics[width=0.5\textwidth]{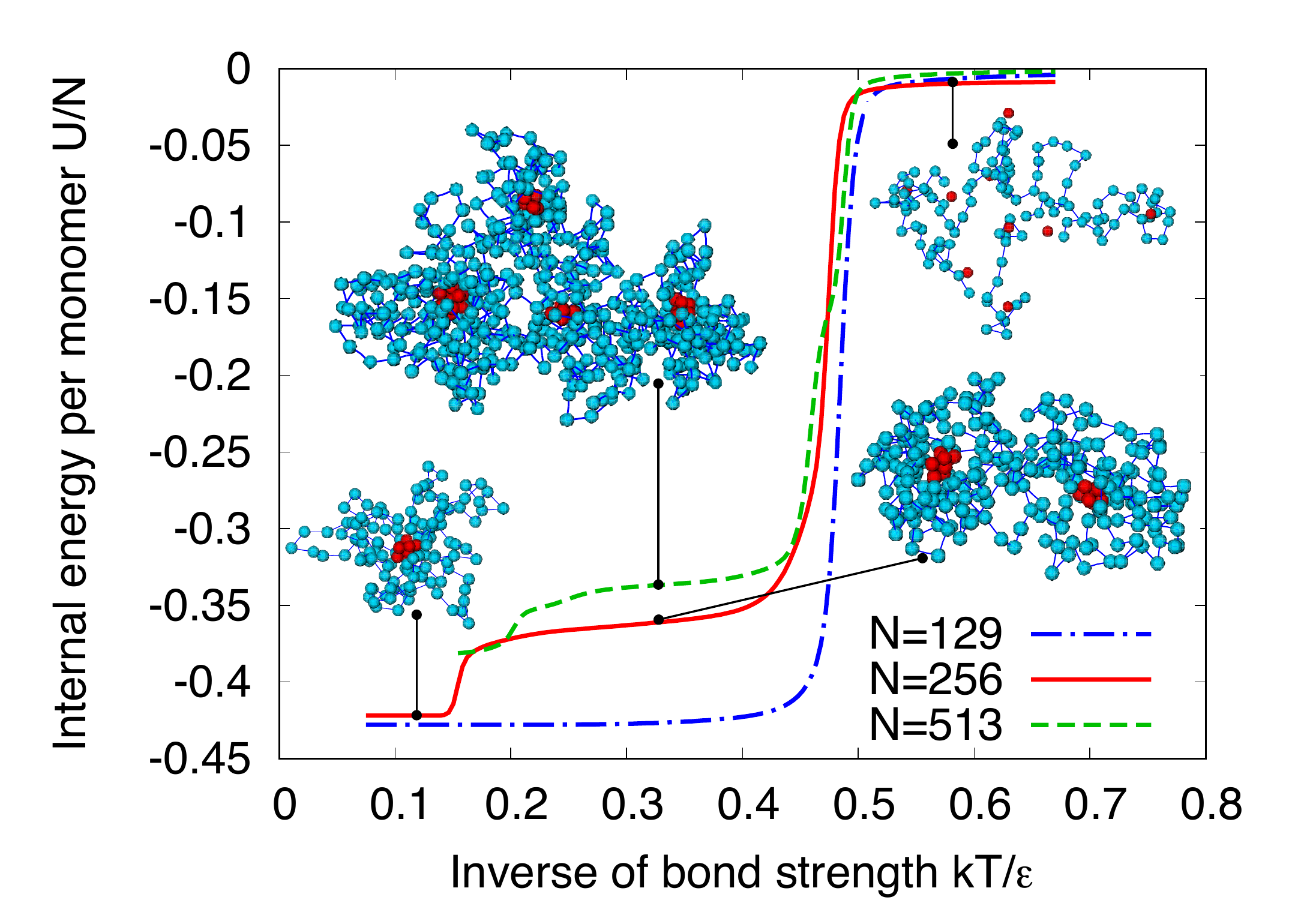}
\caption{ Energy density for co--polymers with an ordered pattern of
  interacting monomers. In these simulations the density of
  interacting monomers $\eta=p/N$ was kept constant to the value
  $1/16$. Simulations were performed with $\varepsilon=2.4$, $R=0.77$,
  $\lambda=1.42$ for $3\times10^{9}$ Monte Carlo sweeps. }
\label{fig:ordered_collapses}
\end{figure}

In the case of equally-spaced bridging points, theoretical arguments support the claim that
multi-domain states are thermodynamically
stable~\cite{scolari2015combined}. To test this hypothesis, we simulated polymers from $N=129$ up to $N=513$ monomers with the parallel-tempring algorithm until the quantities of interest reached convergence, keeping
constant the density of interacting monomers $\eta=p/N$.  As shown in Fig. \ref{fig:ordered_collapses},
polymers with $N=129$ monomers collapse into a single domain, while
the polymers of length $N=256$ and $N=513$ collapse into a
multiple-domain state similar to rosettes. Rosettes are formed by consecutive strands of the chain. In this range of $N$, the number of domains seems to depend linearly
on $p$, as suggested in ref.~\cite{scolari2015combined}. The collapse
for all values on $N$ happens near a temperature of $T\simeq 0.47\varepsilon$
(see Fig. \ref{fig:ordered_collapses}). No phase similar to a random globule, in which the interactions are not correlated with the distance of the interacting monomers along the chain, is observed. All the rosette-like and
multiple-rosette configurations appear to be thermodynamically stable below the coil-globule transition temperature
%
%
(see Supplementary Figure S1).

For longer chains ($N>129$), after a first collapse at higher temperature, the
polymer displays a second collapse at lower temperature from a phase
with higher number of domains to a phase with a lower number of
domains (e.g., see Fig.~\ref{fig:ordered_collapses}, red solid
curve). While the first energy jump displays features similar to a
first-order phase transition, as suggested
in~\cite{scolari2015combined}, the fusion of two domains resembles a
nucleation-like phenomenon, and we speculate that this could be
similar to a second order phase transition.
%
The low-temperature phases are difficult to sample for longer polymers
and thus we could not equilibrate the chain with $N=513$ below
$T=0.14\varepsilon$.  Although we have seen in this range of low
temperatures conformations with three and two rosettes, we are not
able to assess if they are equilibrium states.  Equally, we could not
equilibrate the system at even lower temperatures, at which we expect
the equilibrium state to form a single rosette, because this is
certainly the zero--temperature equilibrium state of the system.

Summing up, our results indicate that new stable multi-domain phases
become available with increasing system size, and that the system can
cross several hierarchical levels of organization with decreasing
number of domains as equilibrium states as the temperature is
decreased, before collapsing into a single domain.

\subsection*{Disorder enhances the stability of multi-domain configurations}

\begin{figure}
\centering
\includegraphics[width=0.5\textwidth]{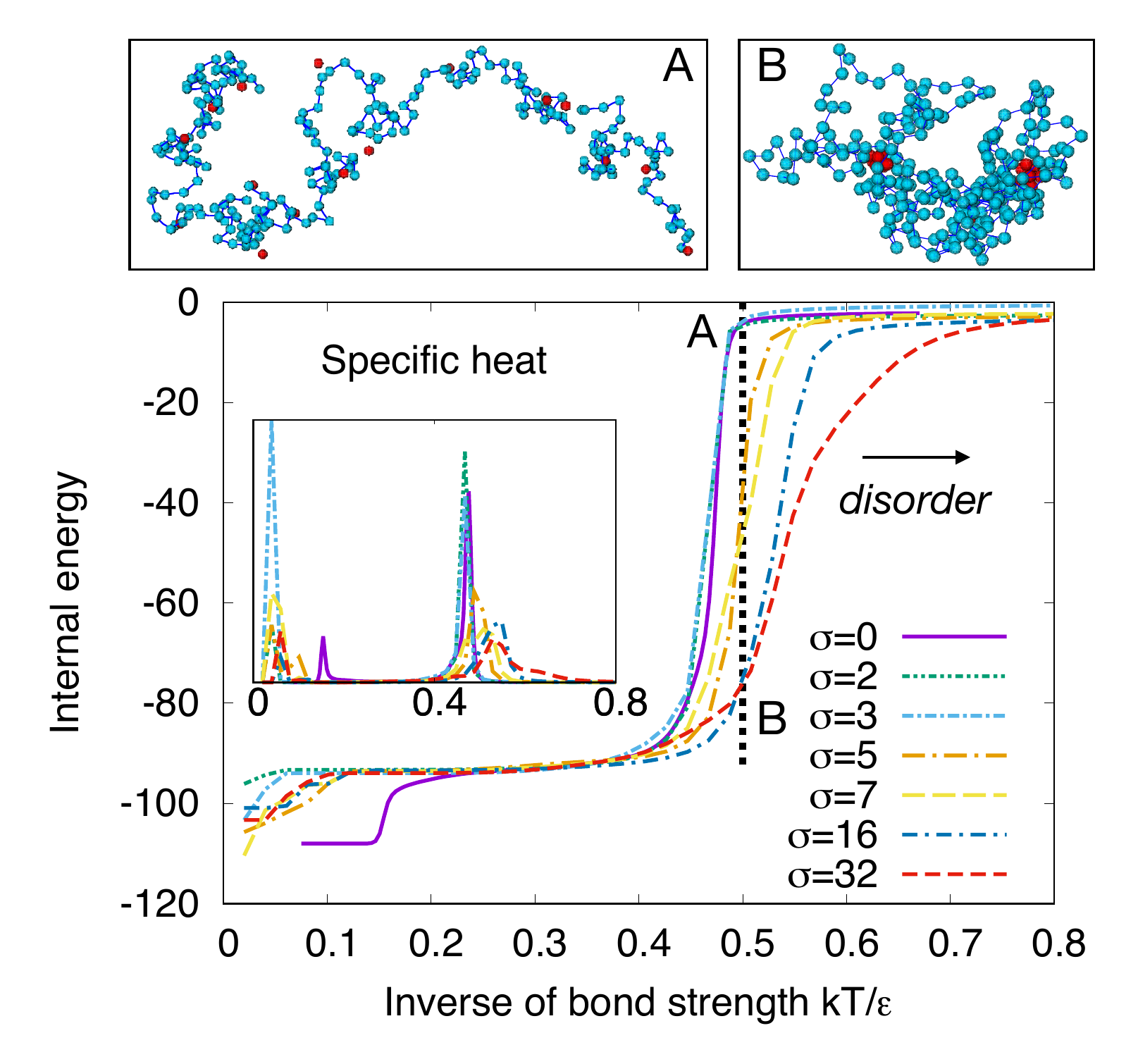}
\caption{Disorder in the positioning of the interacting monomers
  shifts the domain-formation transition towards higher
  temperatures. The plot shows collapse curves of internal energy of a
  polymer of length $N=257$ monomers. Each dashed curve relates to a
  different value of the variance $\sigma$. Two snapshots at the same
  temperature are highlighted comparing the case of regularly-spaced
  interacting monomers (A) to the disordered case (B): while the
  former is clearly in a coil state, the latter appears collapsed into
  a two-domain state.  This is also visible in the specific heat vs
  $kT/\varepsilon$ plot (inset), in which the peak corresponding to
  the transition point smoothens and shifts towards higher
  temperatures in presence of disorder. The simulations were performed
  with $N=257$, $\eta=17/257$, $R=0.77$, $\lambda=1.42$,
  $1.8\cdot10^9$ MC moves. }
\label{fig:disordered_collapses}
\end{figure}

The model with interacting monomers placed every $\eta^{-1}$ other monomers is then extended, introducing a quenched Gaussian displacement of zero mean and variance $\sigma$. 
For $\sigma=0$ we recover the ordered case, while for
$\sigma \gtrsim \eta^{-1}$ we expect a uniform distribution of interacting monomers, not reminiscent of the ordered arrangement.

Before studying the equilibrium properties of the disordered system,
we must show that they are self--averaging, that is that the average
over the disorder is representative of a typical situation. As a rule,
extensive quantities like the internal energy are self--averaging
because of an argument given by Brout \cite{Brout:1959tz}; however
this argument cannot be applied straightforwardly to disordered
polymers, and we checked explicitly in two cases ($\sigma=7$ and
$\sigma=16$, using four realizations of the disorder) that the energy
curves and the number of rosettes do not depend on the specific
realization of the disorder (see Fig. S2 in the Supplementary
Material).

We then performed equilibrium simulations with $\eta^{-1}=16$ and
$\sigma$ varying from 0 to 32 for $N=257$. In all these cases, as
shown in Fig.~\ref{fig:disordered_collapses}, we observe a transition
from a random coil at high temperatures to multi-rosette states. The
transition becomes less sharp with increasing $\sigma$.  Moreover, the
disorder has the unexpected effect of stabilizing the multi-domain
phase, as the transition temperatures become higher. This effect is
accompanied by a broadening of the range of temperatures in which the
multiple domain phase is stable, roughly proportional to $\sigma$.
The inset of Fig.~\ref{fig:disordered_collapses} shows the specific
heat of the system, whose peaks are associated with the transitions.
Two peaks in the specific heat are typically visible in this plot,
corresponding, respectively, to the collapse from coil to two-domain
state (high temperature) and to the transition from two-domain state
to one-domain state (low temperature). These peaks shift apart at
increasing $\sigma$.
The disorder smoothens the collapse curve of the higher transition
only, since the height of the specific-heat peak decreases with
$\sigma$, while it becomes wider.  This suggests that the transition
from coil to multiple-domain states may no longer be switch-like, due
to the emergence of domains of different size at different
temperatures.

\begin{figure}
  \centering \includegraphics[width=0.5\textwidth]{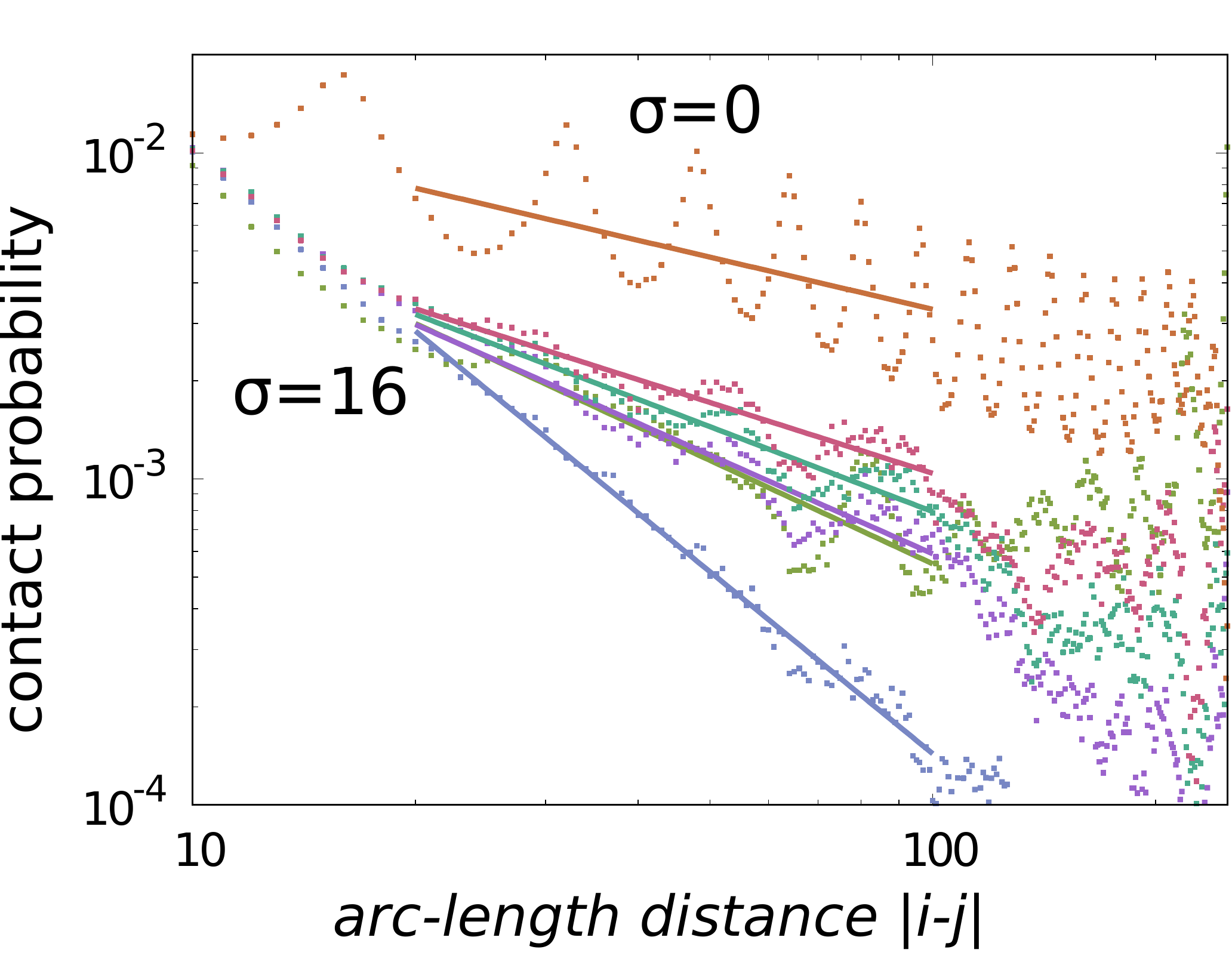}
  \caption{The model shows power-law like scaling of the contact
    probability with arc-length distance.  The plot shows the mean
    logarithm of the contact probability obtained from
    simulations. The average is performed over configurations and over
    all monomers $i$ and $j$, whose inter-monomer distance is
    $|i-j|$. The different curves correspond to the case of ordered
    interacting monomers ($\sigma=0$, orange points) and disordered
    interacting monomers, $\sigma=16$ at temperatures $T=0.49$ (red),
    $T=0.50$ (cyan), $T=0.52$ (green and purple, for two different
    realizations of the disorder), $T=0.53$ (blue). The solid lines
    are linear fits, giving slopes (exponents) 0.53 for $\sigma=0$,
    0.72 for $\sigma=16$ at $T=0.49$, 0.87 at $T=0.50$, 1.01 and 1.05
    at $T=0.52$ and 1.86 at $T=0.53$.}
\label{fig:p}
\end{figure}

We also considered the average contact probability between monomers as
a function of their distance $|i-j|$ along the chain, which is
typically measured from genome contact
maps\cite{imakaev2015modeling,Nicodemi2014}.  Fig.~\ref{fig:p}
compares this function for the case of equally-spaced and disordered
interacting monomers. In the ordered case ($\sigma=0$), the spacing
with the closest interacting induces oscillations in the function, but
the overall trend agrees with a power law with exponent close to 0.5
for values of $|i-j|$ up to distances comparable to $N$ (and therefore
affected by finite-size effects). Disordered chains display exponents
that increase with the temperature between, 0.7 and 1 in the
multi-rosette phase, and up to 1.9 in the coil region (this value is
comparable to the expectation for a self-avoiding chain). The
exponents appear to depend weakly on the specific realization of the
disorder (cf. purple and green points in Fig.~\ref{fig:p}).

\subsection*{Scaling argument for the entropy of a disordered star polymer}

In order give some theoretical support to explain why the
multi-rosette configurations are thermodynamically stable in presence
of disorder, we generalized the scaling argument given in
ref. \cite{scolari2015combined}.  In a
configuration made of $q$ rosettes, each domain has a core made of
$p/q$ monomers and a corona made of $p/q$ loops. Each rosette is
approximated as a star polymer made of $f=p/q$ arms. This description
allows a simple estimate of the entropic contribution of the corona to
the free energy. In absence of disorder the leading term in this
contribution is $f^{3/2}$.  The energetic contribution to the free
energy is the surface tension of each core, which is proportional to
the surface of a single core $(p/q)^{2/3}$ multiplied by the number of
domains. Therefore, the free energy in absence of disorder reads
\begin{equation}
\Delta F \simeq p^{3/2}q^{-1/2}+\varepsilon (p)^{2/3} q^{1/3} \ ,
\end{equation}
which can be minimized with respect to the number of domains $q$, to
find the number of rosettes at equilibrium
\begin{equation}
q_{\mathrm{eq}} \sim  \varepsilon^{-6/5} \ .
\end{equation}

We now estimate how $\Delta F$ changes for disordered distributions of
bridging points in the polymer.  At fixed rosette state, the changes
in the positions of the attractive monomers along the chain do not
affect the energetic term, so we need to compute only the entropic
term for a rosette with loops of random length. To do this, we
approximate the disordered rosette to a star polymer with arms of
random length, and use the blob model for star
polymers~\cite{daoud1982star, witten1986colloid} to describe the
system with a mean-field ansatz. Here we omit intermediate
calculations, which can be found in the supplementary material,
section S1.  To account for the different lengths of the arms, we
impose that the number of arms $f$ is a decreasing function of the
radius,
\begin{equation}
  f(r)=f_{0} \left(\frac{r}{b}\right)^{-\gamma} \ , \label{eq:ansatz}
\end{equation}
where $b$ is the radius of the core of the star and $\gamma \geq 0$.
It is possible to show that this is equivalent to a power-law
distribution of the distance between consecutive attractive monomers
(see supplementary material, section S1, last paragraph).
This assumption does not correspond to the Gaussian displacements of
the bridging points from equally-spaced positions used in our
simulations, and is motivated mainly by the ease of carrying out the
calculation.
%


We can now plug Eq. \ref{eq:ansatz} into a scaling argument similar to
the one found in ref.~\cite{witten1986colloid}.  This calculation
gives a leading term in the entropy that is identical to the one in
absence of disorder,
\begin{equation}
  \Delta F_{\mathrm{entropic}}\simeq f_0 \left[S-\gamma f_{0}^{-1/2}
    \frac{S^2}{2} + \frac{\gamma^2}{4} f_0^{-1} \frac{S^3}{3} \right] 
\end{equation}
with
\begin{displaymath}
  S\simeq \frac{2}{\gamma} f_{0}^{1/2}, 
\end{displaymath}
which implies
\begin{equation}
\Delta F_{\mathrm{entropic}}\simeq (f_0)^{3/2}.
\end{equation}
Thus, this argument supports the existence of stable states in
presence of disorder in the bridging points, and predicts that the
disorder does not change the leading term in the entropy of the
rosettes, and the collapse is qualitatively the same.
Since the leading-order term of the entropy is unaffected in the
extreme power-law spacing between attracting monomers along the chain,
we also expect that this prediction applies for more compact
distributions of the spacing between possible bridging points, such as
the one used in our simulations.
Indeed, we find that the collapsed phase of the polymer of length
$N=257$ exhibits two domains for all values of $\sigma$ we tested,
just as the model in absence of disorder.

In order to rationalize why the simulations show a shift of the
transition towards higher temperatures, which is not predicted by the
above argument, we can notice that the above argument only considers
the star-polymer contribution to the free energy. We can also compare
the typical value of the loop entropy in presence and absence of
disorder, but at fixed $\eta$. In absence of disorder the total entropy of $p$ loops of length $N/p$ is 
\begin{equation}
S_{\mathrm{tot}}\sim p \log(N/p).
\end{equation}
For sufficiently small disorder (i.e. when $\sigma$ is much smaller than $\eta^{-1}$), there are $p$ loops of random length $l_i=|x_i-x_{i+1}|$, where $x_i$ and $x_{i+1}$ are the positions of two consecutive attractive monomers. Since the distribution of $x_i$ is Gaussian, the distribution of $l_i$ is still a Gaussian with mean $\langle l_i \rangle=N/p$. Thus we can compute the total entropy for the system in presence of disorder:   \\
\begin{equation}
S^{\mathrm{dis}}_{\mathrm{tot}}\sim \sum^p_{i=1} \log(l_i).
\label{eq:Sdis}
\end{equation}
We can rewrite eq.~\ref{eq:Sdis} to obtain a relation with $S_{\mathrm{tot}}$:
$$
S^{\mathrm{dis}}_{\mathrm{tot}}\simeq p \sum^p_{i=1} \frac{1}{p}\log(l_i)
$$
$$
\simeq p \langle \log(l_i)\rangle < p \log  \langle l_i \rangle
$$
where in the last line we used the Jensen inequality for concave
functions. This means that
$$
S^{\mathrm{dis}}_{\mathrm{tot}} < S_{\mathrm{tot}}, 
$$
namely that the entropy cost for $p$ loops decreases in the disordered
model, so that the transition temperature increases.
For the same reason, allowing for collapsed states with multiple
domains, one can also speculate that the transition becomes broader
because different regions of the polymer with different local
densities of attractive monomers start to collapse at different
temperatures.


\subsection*{Localization of domains caused by disorder}

\begin{figure*}
  \centering \includegraphics[scale=0.5]{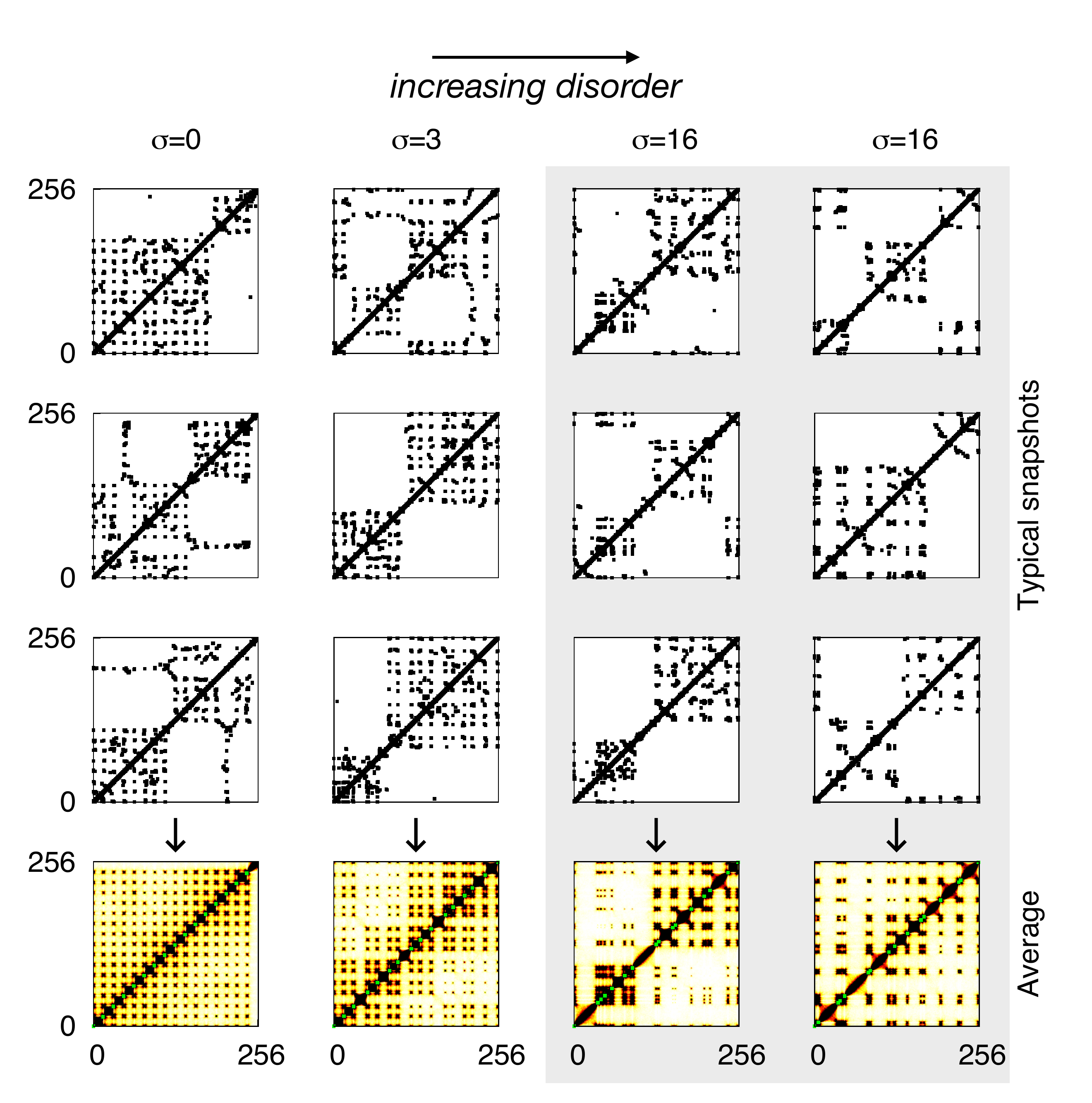}
  \caption{A disordered distribution of interactions can localize the
    domains along the chain. The figure shows contact matrices for
    different conformations of a polymer made of $N=257$ monomers and
    $p=17$ interacting monomers (in green) distant $16$ monomers from
    each other.  The lowest contact maps are the equilibrium average
    of the system. Each column is obtained with a specific realization
    of the disorder, while the two columns with $\sigma=16$ are
    obtained, respectively, with two different realizations of the
    placement of interacting beads.  These simulations were performed
    with $N=257$, $\eta=17/257$, $R=0.77$, $\lambda=1.42$,
    $1.8\cdot10^9$ Monte Carlo sweeps. The replicas used in this image
    are at the temperature $k_BT/\varepsilon=0.3$.}
\label{fig:contact_matrices}
\end{figure*}

In long ordered co--polymers with equally-spaced attracting monomers,
the positions of the domains are invariant for translations along the
chain, and they are free to move along the chain (see
Fig. \ref{fig:contact_matrices}, left column). Different equilibrium
conformations can break this symmetry, displaying domains at specific
positions, but the equilibrium contact map averages out the domains,
re--establishing the translational symmetry (cf. the lowest--left
contact map in Fig.  \ref{fig:contact_matrices}). Only a small effect
due to the finiteness of the chain is observable at the polymer ends;
this would further reduce in longer, more realistic polymers.

Disorder has the effect of localizing the domains, preventing their
averaging out.  The three rightmost columns of
Fig.~\ref{fig:contact_matrices} show the result of simulations
performed with a realization of disorder with $\sigma= 3$ and two
realizations with $\sigma= 16$, choosing $N=257$ and $p=17$. Disorder
breaks the translational symmetry of the system, favouring the
stabilization of domains in specific regions of the chain. As a
consequence, the average map is no longer uniform.
For example, at $\sigma=3$,
contact maps show with high probability a two-blocks structure (figure
\ref{fig:contact_matrices}, second column, bottom panel) that highly
contribute to determine the average map (shown below).

As shown in the case $\sigma=16$ (last two columns of
Fig.~\ref{fig:contact_matrices}), the degree of localization depends
on the specific realization of the disorder.  The figure shows two
contact maps of conformations obtained with two different realizations
of the same distribution of disorder. In the first realization, the
two-block structure has well-defined borders that correspond the
regions around monomer $25$ and $110$. Instead, the second realization
of the same distribution does not show a clear compartmentalization
into two fixed spatial domains, and reallocation of bridging points is
observed around a coarse-grained nearly equally-spaced structure of
organizing centers.
The degree of localization of the domains does not seem to depend
trivially on the organization of the interacting monomers into linear
clusters along the chain (green dots along the diagonal of
Fig. \ref{fig:contact_matrices}). In the case $\sigma=3$, the
displacement from the ordered case is small, but still there is a
higher degree of localization than the case $\sigma=16$ shown in the
rightmost column, where there is a more marked partitioning of
interacting monomers. Thus, the degree of localization appears to
result from a complex balancing between energy and entropy, and cannot
be easily predicted from the location of the interacting monomers.

\section{Discussion and Conclusions}

Our extensive MC simulations give access to the equilibrium properties
of polymers up to up to $N=513$, characterized by a small linear
density of fixed attractive monomers, which can be equally spaced or
disordered. Both in the ordered and disordered case, the phase diagram
of the polymer displays a high-temperature coil phase and a sharp
transition to globular phases with multi-rosette structures. The
states with different number of rosettes are clearly separated from
each other by jumps in the internal energy which resemble first-order
transitions. At highest temperatures we observe states with the
largest number of rosettes, and this number decreases with the
temperature to the one-rosette zero-temperature state. Although the
system size is limited in our simulations by the high computational
cost of equilibrating the system, we can speculate that a hierarchy of
states exists with varying number of rosettes. The maximum observed
number, $n_{\mathrm{max}}\simeq N/128$, is reached just below the
coil-globule transition temperature.
The observed rosettes have the specific feature of involving monomers
that are close along the chain.

The formation of rosette-like domains is a form of microphase
separation (MPS), which the thermodynamic limit is known to take place
in ordered co-polymers and to produce well-defined structures with few
allowed symmetries (lamellar, hexagonal and cubic) in the vicinity of
the homogeneous phase~\cite{Leibler:1980hc}. If disorder is added in
the position of the chemical species, mean-field calculations by de
Gennes show that the MPS phase is suppressed in favour of a glassy
state~\cite{deGennes:1979jp}. Beyond mean field, Shakhnovich and
coworkers showed that fluctuations reduce the glassy temperature
re-establishing the MPS \cite{Sfatos:1993}, by a perturbative approach
in the number of neighbouring monomers common to different
conformations of the chain (disregarded in the mean field). Our
results suggest that correlations between neighbouring monomers play
an important role in defining the phase diagram of this
polymer. Although the size of our polymers is limited by computational
constrains, the rosette-like domains we observed are formed by
consecutive segments of the chain, and suggest that effect of
correlations could be much larger than that suggested by the
perturbative approach. In fact, while the latter predicts a phase
diagram with a second-order transition from a disordered globule to
MPS, we observe what looks like a first-order transition from a random
coil directly to a domain--separated phase. Moreover, at increasing
disorder, the range of temperatures at which MPS occurs increases not
only because the freezing temperature decrease, as predicted in
ref.~\cite{Sfatos:1993}, but also because the high-energy states are
affected (as observed in ref. \cite{Tiana:2011fj}), increasing the
coil--globule transition temperature.

Whether this behaviour is a result of the finiteness of the chain or
is a feature that survives in the thermodynamic limit, we cannot tell
based on our simulations, which are necessarily limited in terms of
the size of the chain. However, the scaling arguments that support the
simulations are not expected to fail in the large--$N$ limit,
suggesting that the phase diagram we propose is stable with respect to
$N$.

An important effect of the disorder is that of localizing the
structural domains in the chain, analogously to what happens with spin
diffusion in the presence of impurities~\cite{Anderson:1958vr}. While
the system, at least in the thermodynamic limit, is invariant for
translation of the domains, and consequently its average equilibrium
contact map is uniform, in presence of quenched disorder the domains
can become localized, resulting in a block equilibrium contact
map. The detailed pattern of blocks, and even how well-defined they
are, does not appear to be a self-averaging quantity, and depends on
the specific positioning of the interacting monomers. These properties
do not seem to be easily predicted from the knowledge of the exact
realization of disorder, in agreement with the general observation
that the identification of the equilibrium states of disordered
systems is a NP-hard problem~\cite{Barahona:1982gj}.

The results obtained with this simple co--polymer model can be useful
to get some insight in the structural organization of
chromosomes~\cite{ringrose2017epigenetics}, which display a
hierarchical set of nested domains~\cite{Zhan:2017kj}.
Little is known about the actual molecular mechanisms responsible for
the formation of domains, at different length scales, in the chromatin
fiber and several models were proposed to account for such an
organization. Some years ago it was suggested that they are the result
of the rapid collapse of the fiber into a non-equilibrium crumpled
globule~\cite{mirny2011fractal}. A model that generates blocks that
are similar to the smallest--scale domains observed in chromatin is
the loop--extrusion, based on the hypothesis that the interaction
between regions of the fiber are mediated by an active, ATP-fueled
protein complex~\cite{Fudenberg2016,Goloborodko2016a}. In other,
equilibrium, models, such as the one we study here, the number of
domains is determined by the number of different interacting
species~\cite{Jost:2014co,Bianco:2017gp}, and xthe formation of
domains is essentially energy-driven.  With the present simple model
we showed that it is not necessary to resort to very complicated
ingredients, but the balance between entropy and energy is enough to
generate stable domains even with a single type of interacting
protein.

Finally, a feature of chromosomes that emerges from experimental data
and that was widely studied in the past years is that the contact
probability between pairs of regions of the same chromosome roughly
scales with their genomic distance with a power law controlled by an
atypical exponent that is variable, but typically lower than 1.5,
behavior that is unexpected for simple homopolymes at
equilibrium~\cite{LiebermanAiden:2009jz,Sanborn:2015dr}. Also in this
case several physical mechanism were proposed
\cite{LiebermanAiden:2009jz,Barbieri:2012iw,Zhan:2016ds,Fudenberg2016}. Our
results suggest that even a simple model as the one we propose here
produces equilibrium contact probability functions that can be fitted
with power laws of genomic distance, with exponents that are lower
than those of homopolymers, and in overall agreement with the trends
of experimental data. In our model, the slopes of this contact
probability depend on the disorder strength and on the stabilization
energy of the domains.




\begin{thebibliography}{99}

\bibitem{imakaev2015modeling}
M.~V. Imakaev, G.~Fudenberg, and L.~A. Mirny, ``Modeling chromosomes: Beyond
  pretty pictures,'' {\em FEBS letters}, vol.~589, no.~20PartA, pp.~3031--3036,
  2015.

\bibitem{Lagomarsino2015}
M.~Cosentino~Lagomarsino, O.~Esp{\'{e}}li, and I.~Junier, ``From structure to
  function of bacterial chromosomes: Evolutionary perspectives and ideas for
  new experiments.,'' {\em FEBS Lett}, vol.~589, pp.~2996--3004, Oct 2015.

\bibitem{Nicodemi2014}
M.~Nicodemi and A.~Pombo, ``Models of chromosome structure,'' {\em Current
  opinion in cell biology}, vol.~28, pp.~90--95, 2014.

\bibitem{Dame2011}
R.~T. Dame, O.~J. Kalmykowa, and D.~C. Grainger, ``Chromosomal macrodomains and
  associated proteins: implications for dna organization and replication in
  gram negative bacteria.,'' {\em PLoS Genet}, vol.~7, p.~e1002123, Jun 2011.

\bibitem{schwarzer2017two}
W.~Schwarzer, N.~Abdennur, A.~Goloborodko, A.~Pekowska, G.~Fudenberg,
  Y.~Loe-Mie, N.~A. Fonseca, W.~Huber, C.~H. Haering, L.~Mirny, {\em et~al.},
  ``Two independent modes of chromatin organization revealed by cohesin
  removal,'' {\em Nature}, vol.~551, no.~7678, p.~51, 2017.

\bibitem{Fudenberg2016}
G.~Fudenberg, M.~Imakaev, C.~Lu, A.~Goloborodko, N.~Abdennur, and L.~A. Mirny,
  ``Formation of chromosomal domains by loop extrusion,'' {\em Cell reports},
  vol.~15, no.~9, pp.~2038--2049, 2016.

\bibitem{Goloborodko2016a}
A.~Goloborodko, J.~F. Marko, and L.~A. Mirny, ``Chromosome compaction by active
  loop extrusion,'' {\em Biophysical journal}, vol.~110, no.~10,
  pp.~2162--2168, 2016.

\bibitem{pant2004mutation}
V.~Pant, S.~Kurukuti, E.~Pugacheva, S.~Shamsuddin, P.~Mariano, R.~Renkawitz,
  E.~Klenova, V.~Lobanenkov, and R.~Ohlsson, ``Mutation of a single ctcf target
  site within the h19 imprinting control region leads to loss of igf2
  imprinting and complex patterns of de novo methylation upon maternal
  inheritance,'' {\em Molecular and cellular biology}, vol.~24, no.~8,
  pp.~3497--3504, 2004.

\bibitem{RN49}
M.~C. Noom, W.~W. Navarre, T.~Oshima, G.~J. Wuite, and R.~T. Dame, ``H-ns
  promotes looped domain formation in the bacterial chromosome,'' {\em Curr
  Biol}, vol.~17, no.~21, pp.~R913--4, 2007.

\bibitem{brackley2016simulated}
C.~A. Brackley, J.~Johnson, S.~Kelly, P.~R. Cook, and D.~Marenduzzo,
  ``Simulated binding of transcription factors to active and inactive regions
  folds human chromosomes into loops, rosettes and topological domains,'' {\em
  Nucleic acids research}, vol.~44, no.~8, pp.~3503--3512, 2016.

\bibitem{nazarov2015statistical}
L.~I. Nazarov, M.~V. Tamm, V.~A. Avetisov, and S.~K. Nechaev, ``A statistical
  model of intra-chromosome contact maps,'' {\em Soft Matter}, vol.~11, no.~5,
  pp.~1019--1025, 2015.

\bibitem{junier2010spatial}
I.~Junier, O.~Martin, and F.~K{\'e}p{\`e}s, ``Spatial and topological
  organization of dna chains induced by gene co-localization,'' {\em PLoS
  computational biology}, vol.~6, no.~2, p.~e1000678, 2010.

\bibitem{scolari2015combined}
V.~F. Scolari and M.~C. Lagomarsino, ``Combined collapse by bridging and
  self-adhesion in a prototypical polymer model inspired by the bacterial
  nucleoid,'' {\em Soft matter}, vol.~11, no.~9, pp.~1677--1687, 2015.

\bibitem{tiana2015montegrappa}
G.~Tiana, F.~Villa, Y.~Zhan, R.~Capelli, C.~Paissoni, P.~Sormanni, E.~Heard,
  L.~Giorgetti, and R.~Meloni, ``Montegrappa: an iterative monte carlo program
  to optimize biomolecular potentials in simplified models,'' {\em Computer
  Physics Communications}, vol.~186, pp.~93--104, 2015.

\bibitem{swendsen1986replica}
R.~H. Swendsen and J.-S. Wang, ``Replica monte carlo simulation of
  spin-glasses,'' {\em Physical Review Letters}, vol.~57, no.~21, p.~2607,
  1986.

\bibitem{ferrenberg1989optimized}
A.~M. Ferrenberg and R.~H. Swendsen, ``Optimized monte carlo data analysis,''
  {\em Physical Review Letters}, vol.~63, no.~12, p.~1195, 1989.

\bibitem{Brout:1959tz}
R.~Brout, ``{Phys. Rev. 115, 824 (1959) - Statistical Mechanical Theory of a
  Random Ferromagnetic System},'' {\em Physical Review}, 1959.

\bibitem{daoud1982star}
M.~Daoud and J.~Cotton, ``Star shaped polymers: a model for the conformation
  and its concentration dependence,'' {\em Journal de Physique}, vol.~43,
  no.~3, pp.~531--538, 1982.

\bibitem{witten1986colloid}
T.~Witten and P.~Pincus, ``Colloid stabilization by long grafted polymers,''
  {\em Macromolecules}, vol.~19, no.~10, pp.~2509--2513, 1986.

\bibitem{Leibler:1980hc}
L.~Leibler, ``{Theory of Microphase Separation in Block Copolymers},'' {\em
  Macromolecules}, vol.~13, pp.~1602--1617, Nov. 1980.

\bibitem{deGennes:1979jp}
P.~G. de~Gennes, ``{Theory of long-range correlations in polymer melts},'' {\em
  Faraday Discussions of the Chemical Society}, vol.~68, pp.~96--103, Jan.
  1979.

\bibitem{Sfatos:1993}
C.~D. Sfatos, A.~M. Gutin, and E.~I. Shakhnovich, ``{Phase diagram of random
  copolymers},'' vol.~48, no.~1, pp.~465--475, 1993.

\bibitem{Tiana:2011fj}
G.~Tiana and L.~Sutto, ``{Equilibrium properties of realistic random
  heteropolymers and their relevance for globular and naturally unfolded
  proteins.},'' vol.~84, p.~061910, Dec. 2011.

\bibitem{Anderson:1958vr}
P.~W. Anderson, ``{Absence of Diffusion in Certain Random Lattices},'' {\em
  Phys. Rev.}, vol.~109, pp.~1492--1505, 1958.

\bibitem{Barahona:1982gj}
F.~Barahona, ``{On the computational complexity of Ising spin glass models},''
  {\em Journal of Physics A: Mathematical and General}, vol.~15,
  pp.~3241--3253, Oct. 1982.

\bibitem{ringrose2017epigenetics}
L.~Ringrose, {\em Epigenetics and Systems Biology}.
\newblock Academic Press, 2017.

\bibitem{Zhan:2017kj}
Y.~Zhan, L.~Mariani, I.~Barozzi, E.~G. Schulz, N.~Bl{\"u}thgen, M.~Stadler,
  G.~Tiana, and L.~Giorgetti, ``{Reciprocal insulation analysis of Hi-C data
  shows that TADs represent a functionally but not structurally privileged
  scale in the hierarchical folding of chromosomes},'' {\em Genome Research},
  vol.~27, pp.~479--490, Mar. 2017.

\bibitem{mirny2011fractal}
L.~A. Mirny, ``The fractal globule as a model of chromatin architecture in the
  cell,'' {\em Chromosome research}, vol.~19, no.~1, pp.~37--51, 2011.

\bibitem{Jost:2014co}
D.~Jost, P.~Carrivain, G.~Cavalli, and C.~Vaillant, ``{Modeling epigenome
  folding: formation and dynamics of topologically associated chromatin
  domains.},'' {\em Nucleic Acids Research}, vol.~42, pp.~9553--9561, Sept.
  2014.

\bibitem{Bianco:2017gp}
S.~Bianco, A.~M. Chiariello, C.~Annunziatella, A.~Esposito, and M.~Nicodemi,
  ``{Predicting chromatin architecture from models of polymer physics.},'' {\em
  Chromosome research : an international journal on the molecular,
  supramolecular and evolutionary aspects of chromosome biology}, vol.~25,
  pp.~25--34, Mar. 2017.

\bibitem{LiebermanAiden:2009jz}
E.~Lieberman-Aiden, N.~L. van Berkum, L.~Williams, M.~Imakaev, T.~Ragoczy,
  A.~Telling, I.~Amit, B.~R. Lajoie, P.~J. Sabo, M.~O. Dorschner, R.~Sandstrom,
  B.~Bernstein, M.~A. Bender, M.~Groudine, A.~Gnirke, J.~Stamatoyannopoulos,
  L.~A. Mirny, E.~S. Lander, and J.~Dekker, ``{Comprehensive mapping of
  long-range interactions reveals folding principles of the human genome},''
  {\em Science}, vol.~326, pp.~289--293, Oct. 2009.

\bibitem{Sanborn:2015dr}
A.~L. Sanborn, S.~S.~P. Rao, S.-C. Huang, N.~C. Durand, M.~H. Huntley, A.~I.
  Jewett, I.~D. Bochkov, D.~Chinnappan, A.~Cutkosky, J.~Li, K.~P. Geeting,
  A.~Gnirke, A.~Melnikov, D.~McKenna, E.~K. Stamenova, E.~S. Lander, and E.~L.
  Aiden, ``{Chromatin extrusion explains key features of loop and domain
  formation in wild-type and engineered genomes},'' {\em Proceedings of the
  National Academy of Sciences of the United States of America}, vol.~112,
  pp.~E6456--E6465, Nov. 2015.

\bibitem{Barbieri:2012iw}
M.~Barbieri, M.~Chotalia, J.~Fraser, L.-M. Lavitas, J.~Dostie, A.~Pombo, and
  M.~Nicodemi, ``{Complexity of chromatin folding is captured by the strings
  and binders switch model.},'' {\em Proceedings of the National Academy of
  Sciences of the United States of America}, vol.~109, pp.~16173--16178, Oct.
  2012.

\bibitem{Zhan:2016ds}
Y.~Zhan, L.~Giorgetti, and G.~Tiana, ``{Looping probability of random
  heteropolymers helps to understand the scaling properties of biopolymers.},''
  vol.~94, pp.~032402--10, Sept. 2016.

\end{thebibliography}
\end{document}